# Planar alignment of Liquid Crystals on h-BN for Near-Eye Displays

M. Arslan Shehzad,[1,2,3*] Kanghee Won,[4] Yongho Seo,[3] Xinqi Chen[1,2]

[1]Department of Material Science and Engineering, Northwestern University, Evanston, Illinois 60208, United States

[2]Northwestern University Atomic and Nanoscale Characterization Experimental (NU*ANCE*) Center, Northwestern University, Evanston, Illinois 60208, United States

[3]Department of Nanotechnology and Advance Material Engineering, Sejong University, Seoul 05006, South Korea

[4]Samsung Advanced Institute of Technology (SAIT), Samsung Electronics, Co. Ltd., Suwon-si, Gyeonggido 16678, South Korea

**Corresponding Author:** arslan@northwestern.edu

## Abstract:

Liquid crystals alignment on 2D materials is known due to their intrinsic van der Waals interaction. Here, we provide a proof of concept where alignment of liquid crystals (LC) on 2D materials can be utilized to control incident light. We demonstrate that direction and strength of electric field can tune the alignment of LC adsorbed on 2D surface. Due to degeneracy of alignment LC on hexagonal surface, it prefers to align in three multiple states. It is shown that electric field can reorient the direction of liquid crystal alignment on Zigzag and armchair edges of h-BN. The ab initio calculations confirmed the favorable adsorption configurations of liquid crystal molecule on hexagonal boron nitride surface. This concept provides a pathway towards dynamic high-quality pixels with low power consumption and could define a new avenue in near-eye displays.

## Introduction:

Displays have become an essential part of our life in the form of mobile phones, desktop computers, monitors, projectors and so on. Liquid crystal display (LCD) in its conventional form requires a backup light source which limits its form factor. Interaction of LC molecules is an important element which can be used to determine the alignment of these crystals.[1] This alignment can be tuned by using different surface morphologies at the interface and the external electric field.[2, 3] Alignment is one of the critical factor to determining the attributes of displays. Moreover, this alignment on smooth surface can only be two-dimensional or homeotropic.[2, 4] Previously most of the alignment and controllability is done using rubbed polymer films,[5] photopolymerization,[6] and predefined layers.[7] However, most of these methods require surface anisotropy and it is challenging to control these structures at the molecular level. Nematic LC due to their soft nature have the tendency to orient spontaneously, via electric field.[8, 9] For LC non-volatile device it is an urgent need to have multistable states with color tunability.[10] [11] Moreover, precise alignment of LC can help to overcome the challenges in ultra-thin flat displays.[12]

Controllability of alignment is another key aspect, LC devices with multistable orientations can help to extend the functionality of different colors.[8] Utilizing the birefringence property and alignment of LC on 2D surfaces help to visualize grains and boundaries of 2D monolayers.[13-15] Previously,[14] we confirmed that LC molecules prefers to align along the crystal orientation of hexagonal 2D structures.[14, 16, 17] Similar technique is employed to determine defects in CVD-growth of 2D layers i.e. graphene, $MoS_2$.[18] While electrically actuated ultrathin optics would advance the development of compact imaging systems, however, LC-based varifocal alignment has remained unintended so far. Moreover, controllability of ultrafine alignment, fast switching and wave control are the major roadblocks to utilize LC in AR/VR and near eye displays.[12, 19] In-plane alignment as shown via 2D surfaces can be used to overcome these challenges. [20, 21]

In this work, we report alignment of LC molecules on CVD grown hBN. We proposed an advanced method to comprehend multistable states on a single grain of hexagonal surface. We demonstrate that external electric force can be used to control the orientation of LC-domain. Furthermore, it is confirmed that both Zigzag and armchair alignment can be used to align. Multiple degenerate states with different of colors can be achieved by varying the combination of the electric field and the intensity of applied field. We are certain that this approach helps to develop non-volatile and flexible displays and could also improve the utility of 2D Nanomaterials in displays arena.

**Results and Discussion:**

Hexagonal Boron nitride is grown and transferred on the silicon wafer ($SiO_2$) as shown in Figure S1. (Details are in the experimental section) The liquid crystal director profile in our device is essential for its proper utilization and optimization and it is investigated using a liquid crystal modeling simulation program. Dielectric medium with dielectric constant equal to h-BN is utilized in the overall scheme. Similarly, both top and bottom surfaces are randomly aligned without any preferential orientation. Figure 1 (b) shows the field profile the boundary condition of one electrode is set to ground (0V) whereas the other two electrodes is set to sweep mode voltage, which changed the voltage from 0V to 10V and performed the calculations at each voltage. Considering the liquid crystal molecules on top of insulating surface, director configuration depends on the interaction between elastic forces, external field, and surface treatment. Figure 1(c) show the LC filled top-down view of the device. Upon applying electric field on the left electrode as shown in Figure 1(e), the liquid crystal molecules follow the direction of the electric field for the case of a positive dielectric anisotropy. The orientation of the liquid crystal director is defined by the tilt angle θ and azimuthal angle φ. It is assumed that the tilt angle is very low (~2°) and the azimuthal angle is randomly distributed between 0~360° within the simulation. However, the azimuthal angle is randomly distributed at an interval of 60° due to the hexagonal symmetry of our BN surface in the actual device. Interestingly, it is observed that after electric field pulse (e), LC molecules retain their new alignment state without any backup force. This theoretically confirms that LC alignment can be controlled on the dielectric surface.

To measure the effect of uniform electric field effect, two electrodes device is fabricated, as shown in Figure 2. The cell gap is adjusted to 1μm and is filled with Liquid crystals (5CB Sigma Aldrich). DC pulses of 2.5 ms width with different polarities are applied to tune the preferential states. (a, d) A grain, outlined on the figure with a dotted red line, initially showed greenish, like the background color. (b, e) When pulses (10, -10V) are applied on two electrodes, respectively, the domain became blue confirming the change of alignment. Inset electric field simulation profile which confirms the direction of field from left to right. Later the color remains stable without the any external force which confirms the stability of alignment. (c, f) Upon applying reversed polarity (-10, 10V), the domain color changed to red, and the changed color remained consistent without any external force. Inset show the direction of field from right to left confirming the configuration. The whole process is repeatable in the different cycles of the pulse sequence. These experimental results can be explained by reorientation of the LC director on the grain. (g) The alignment of the LC molecules is roughly perpendicular to the direction of electric field, since the rubbing of the top layer is in same direction. (h) When external field is applied from the left to the right electrode, LC directors near the h-BN surface is rotated +60º toward the other preferential state, causing the LC layer to be twisted. (h) Contrary to this if the direction of field is applied opposite, the LC directors is rotated -60º toward the third preferential state, causing twisting in the opposite direction as shown in ON state of Figure 2(h).

Similarly, to confirm this argument and further control the alignment on individual domain, three electrodes-based device is fabricated. Figure 3(a-c) show schematics of the fabricated device where three different electric field direction is used to control the alignment of the liquid crystals. Considering the degeneracy of alignment of hexagonal structure, three different alignment states are feasible as shown in Figure 3(c). By carefully controlling the field direction and strength three different states of alignment can be achieved. Two complete domains are intentionally encapsulated in-between electrodes as shown in Figure 3(e-g). 50ms pulse is applied in a way that positive pulse was applied to first electrode and immediately after this negative pulse was applied to the other electrodes to get inhomogeneous field profile as shown in graph of (e) where E1, E2 and E3 are 1st, 2nd and 3rd electrodes. (e) In-situ POM observation (scale bar is 10µm) show; when first sequence is applied, electric field (Red) lines moving away from electrode (E1) will align LC along the domain "a" on any of the three favored orientation of hexagonal surface. Inset shows the diffraction pattern obtained on reflected mode to see the change. Blue color is evident on domain "a" and red for domain "b". (f) similarly, second sequence is applied, and inhomogeneous field profile is obtained from E2 to E1 and E3. Inset diffraction of laser was diffused confirming the change in the orientation of underlying surface. Interestingly it was found that domain "a" was tuned from blue to green without any effect on domain "b" even though the field was applied to it. This confirms the selective controllability of domains. Similarly, when field profile was from E3 to E2 and E1 domain "b" was tuned from red to green while domain "a" was tuned back to its original blue color. It is established that inhomogeneous field could be exploited to selectively regulate the alignment on different domains.

To estimate the threshold voltage $V_{th}$, positive pulses are applied to the right electrode and the other electrodes are grounded in Figure 4. Voltage is increased linearly to 10 V, the grain outlined with a dotted white line is witnessed. It was observed that the grain of red color originally was converted to green approximately at 6 V as shown in Figure 4 (a). Thus, this voltage can be the threshold value for LC domain reorientation. Even though a preliminary slow change before 6 V is found, it is attributable to surface contamination due to polymer residue on top of the h-BN. At this threshold voltage, the electric force from the external field becomes equal to the anchoring force holding the LC to the initial orientation ($\varphi_i = 0$), and the directors start jumping to the next preferential orientation ($\varphi_f = 60°$). The same voltage sequence and pulses is applied to the other electrode as shown in Figure 4 (b). Similarly, the blue domain of a grain is converted to green with almost similar threshold voltage ($V_{th}$ = 6 V). Considering the mean distance between electrodes to be 10 µm, the threshold electric field intensity $E_{th} = 0.6$ V/µm is estimated, which is much lower than that of bistable LC on four-fold symmetry patterned polyimide reported by Kim, et al.[22] The adjacent graph in Figure 4 (c, d) shows the changeover of the intensities of the colors, which were estimated from average values of all grain pixels from the digitized image. This measurement is repeated to determine the threshold voltage on another sample with different shape of the grain, and similar threshold field was obtained.

As seen above the azimuthal strength of the LC/h-BN interface can be estimated from the threshold voltage. [22] The LC layers can be represented by the azimuthal angle of the director, φ, which is a function of the distance from the surface. The anchoring energy is mainly determined by the azimuthal angle difference $\Delta\varphi = \varphi_f - \varphi_i$ at the LC/h-BN. This energy can be expressed by [23, 24] $W(\varphi_f, \varphi_i) = (W_o/2)\sin^2(\Delta\varphi)$, where $W_o$ is the anchoring strength. In case of hexagonal surface, the anchoring energy will have a maximum value i.e., $\Delta\varphi = \pi/6$. Upon applying the external force via electric field due to distinct π/6 states, the LC jumps to a next preferential orientation on hexagonal surface, which ultimately change the wavelength of incident light.

To further confirm the effect of electric field on alignment of LC DFT calculations are used. We checked external field effect on the orientation of a 5CB molecule on an h-BN nanoribbon (width: ~2 nm) in the zigzag and the armchair direction as shown in Figure 5. The two optimized structures are obtained in the absence of external electric field, and then a uniform external electric field is applied to the structure. As we know the, the adsorption energy of the 5CB molecule along the armchair direction is stronger than that in the zig-zag direction in the absence of applied electric field as shown in Figure S2, 3. However, when applying the external electric field, the orientation of the 5CB molecule changes from armchair to the zig-zag direction with electric field intensity of more than 0.03 V/Å as shown in the Figure (b). This confirms that alignment of LC when placed in an external electric field, an external electric dipole moment tends to be aligned parallel to the field to make potential energy lower. As this simulation is based on single LC molecule and nanoscale finite size substrate, the threshold electric field is estimated to be much stronger than the experimental result with collective LC dipole array which confirms our hypothesis.

**Conclusion:**

In this work, we discussed an innovative method for realizing multistable states of LC molecules on CVD grown h-BN. Planar electric field is used to control the alignment of the liquid crystal. Experimentally, several states with a variety of colors are confirmed by varying the direction and strength of electric field. From electro-optical measurements, we proposed further possibilities such as a novel display of various colors on a single pixel. Overall, the new design paradigm presented in this work provides a substitute approach toward the realization of filter-free in-plane nonvolatile displays without both a polarizer and a backlight unit. Owing to the molecular alignment symmetry of the LC on the hexagonal lattice, we claim that an ultra-high-resolution display with a molecular-scale pixel can be realized.

**Experimental Section:**

CVD-grown h-BN is transferred onto a silicon substrate using a conventional wet chemical route. LC (Sigma Aldrich5CB) was spin coated to be 500nm thick and heated above its isotropic temperature (60 °C) and allowed to cool at a slow rate of 1 °Cmin$^{-1}$ to minimize the effect of thermal flow. A 0.5 µl liquid crystal was spin coated (3000rpm) for 60s on CVD-grown boron nitride to cover the surface. The LC molecules are directly aligned along the surface of the h-BN and domain orientation/boundaries can be observed under POM with cross polarizers. An LC cell is fabricated with a rubbed layer on the top electrode and polyvinyl alcohol (PVA) is spin coated on a glass slide. The film is uni-directionally rubbed and placed over the LC-coated surface to confine the anchoring orientation at the top layer. This gap is controlled by polystyrene beads to obtain a 1–2 µm thick LC cell.

The LC alignment, boundaries and domain orientation were observed using a polarizer optical microscope (BX-51, Olympus Corp.). The axis of the polarizer was adjusted to be perpendicular to the analyzer.

Optical lithography with a lift-off process was employed using a positive photoresist to pattern the different electrodes. To investigate this electrical pulse response of aligned LC molecules, a digital signal generator is used which is equipped with a digital-to-analog data acquisition interface card (National Instrument's PCI Express 6353). National Instruments' Data Measurement Services Software (NI-DAQmx) with LabVIEW is used to program pulse voltage sequences with a period of 2.5ms.

Density Functional theory is employed to using the projector-augmented wave pseudopotential implemented in Vienna ab initio simulation package (VASP) code with the cut-off energy of 500 eV. For exchange-correlation functional, the generalized gradient approximation (GGA) proposed by Perdew, Burke, and Ernzerhof is used. We also included the DFT-D2 method for the van der Waals correction. The $6 \times 8$ h-BN supercell of an orthorhombic primitive cell was adopted as a substrate for liquid crystal molecules.

**Acknowledgments:** MAS and XC would like to acknowledge KECK II, EPIC facility of Northwestern University's NUANCE Center, which has received support from the SHyNE Resource [NSF Grant ECCS-2025633], Northwestern's MRSEC program (NSF Grant DMR-1720139), the Keck Foundation, and the State of Illinois through IIN.

**Notes:** Authors declare no competing Financial Interest

**Figures:**

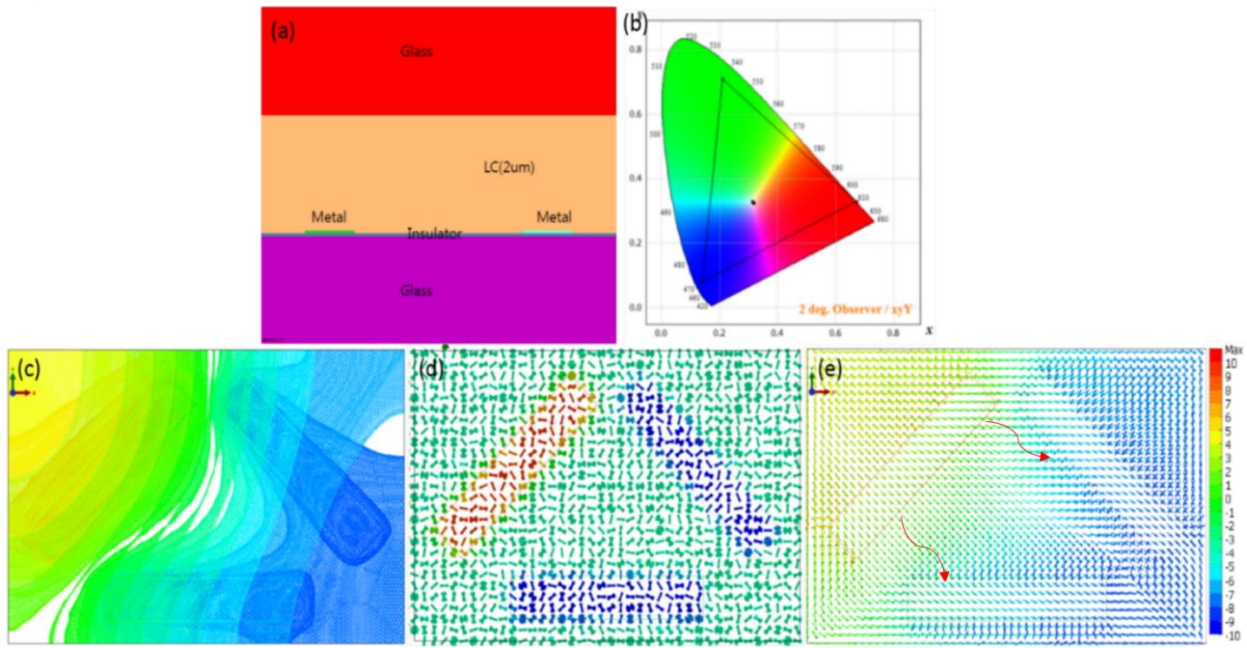

**Figure 1: Simulation of LC director profile.** (a) shows the schematics of device with metal electrodes on an insulating surface. (c) show the LC filled top-down view of the device. (d) shows upon applying 50 msec of voltage pulse due to the impact of field distribution LC molecules change their orientation. (e) Upon applying directional pulse from one electrode to other, it clearly shows the directional change of LC orientation.

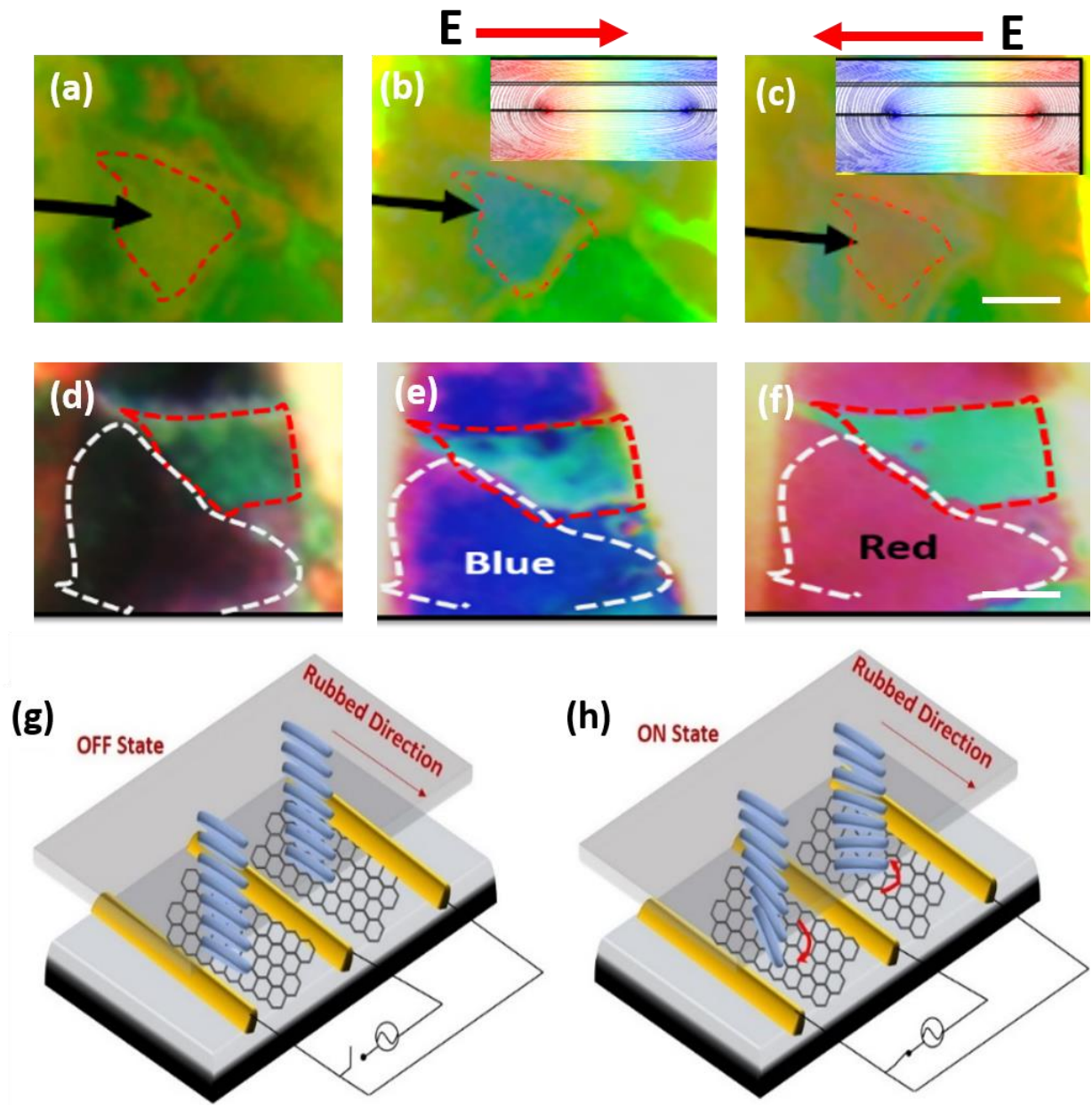

**Figure 2: Homogenous Field effect. (a-f)** show two different domains which are controlled by uniform electric field in between two electrodes. (a) Before electric field liquid crystal can align on any of the preferential however upon applying the electric field, domains changed their alignment as shown in (b, e) and (c, f). Possible phenomena are shown in figure (g) and (h) where upon applying E direction of field help to tune the alignment of LC on 2D layer. Scale bar is 5 μm.

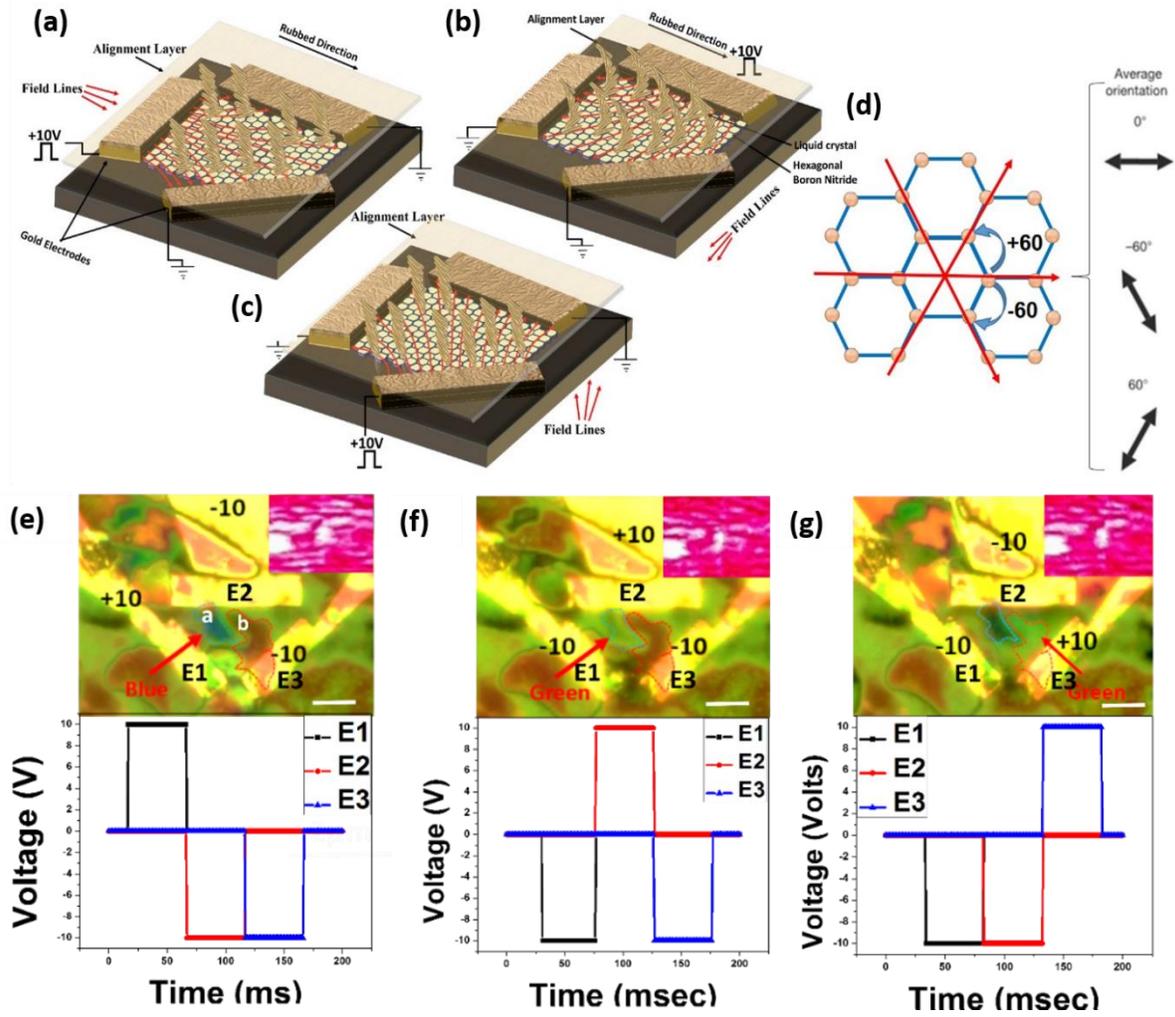

**Figure 3: Inhomogeneous field to selectively control the alignment.** Three electrode device is fabricated as shown in schematics (a-c) where electric field is applied to specific Electrode. Due to degeneracy of alignment on hexagonal surface it has the freedom to align at three different orientations shown in (d). (e) shows two different domains with different colors confirming difference in alignment are used. (f) When field is applied to E2, domain "a" is changed from blue to red confirming the change in the alignment. Similarly, (g) show the opposite alignment where domain "b" is controlled by changing the direction of field. Scale bar represents 1μm.

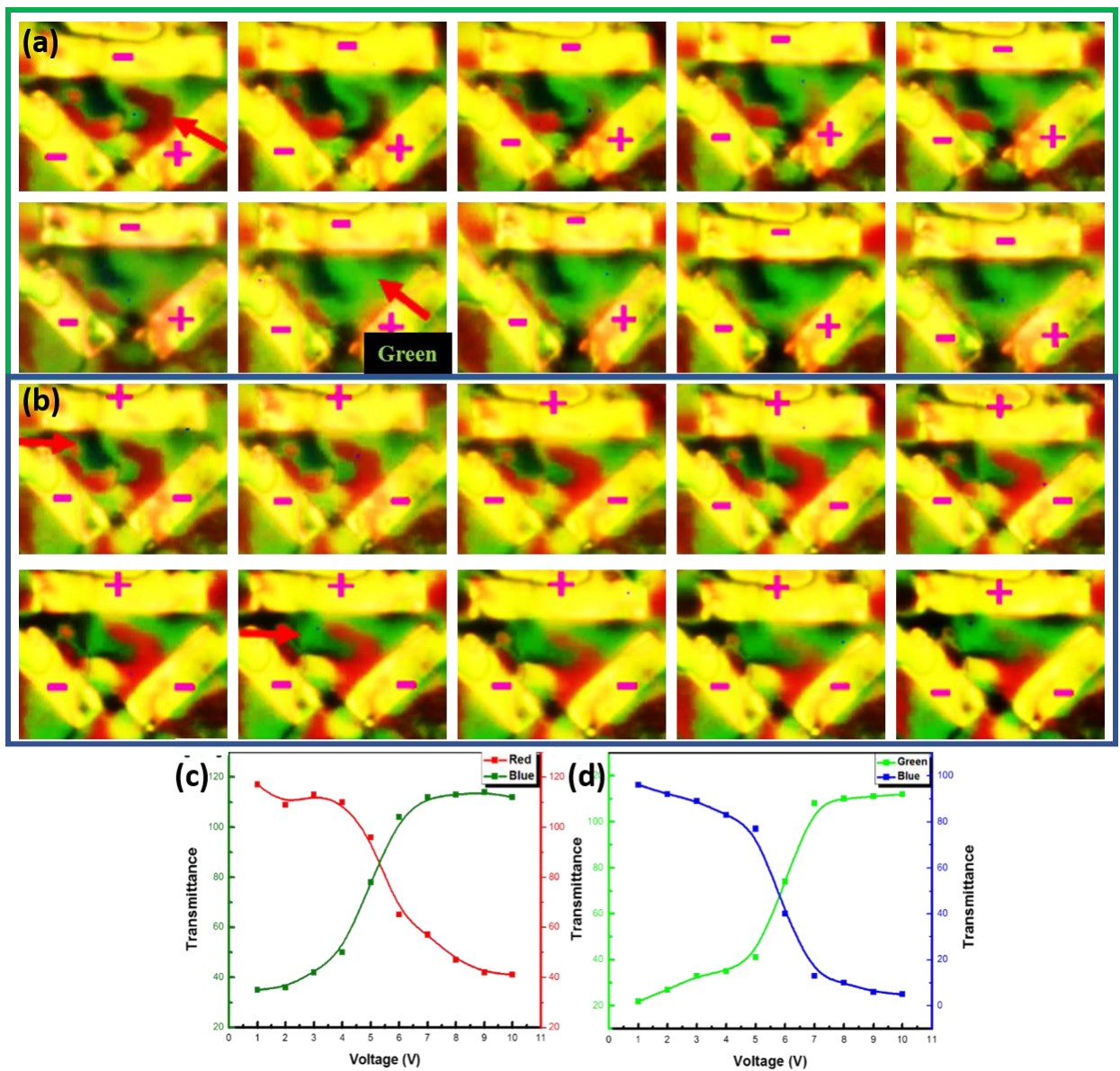

**Figure 4: Effect of strength of Electric field:** To play with the strength of electric field to determine the threshold voltage, two domains as shown in figure 3 are used. Voltage was biased from 1 to 10V respectively as shown in (a) and (b). Both shows that the specific domain alignment was changed at 6V. Graphs in (c, d) depicts the transition of red to green in domain "a" and blue to green in domain "b".

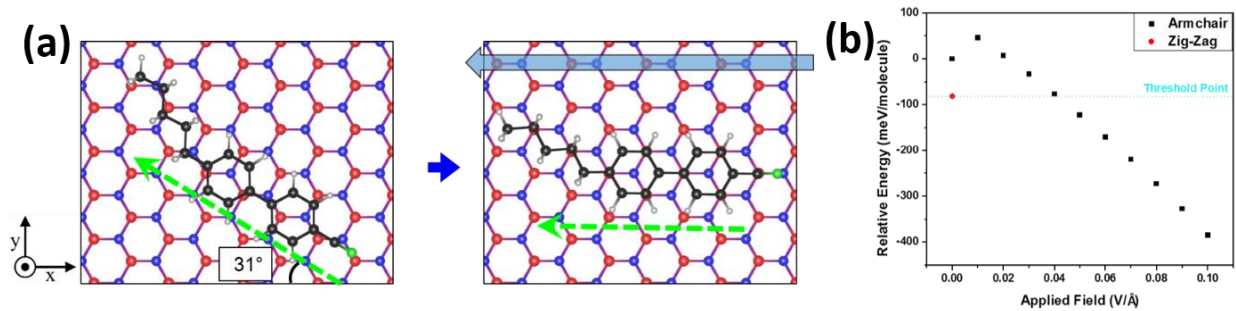

**Figure 5: External field effect on the orientation of a 5CB molecule the h-BN sheet in the zigzag (left) and the armchair direction (right).** These two structures were obtained using ionic relaxation calculations (atoms allowed to move) in the absence of external electric field. When applied a uniform external electric field (E-field) to the structure in the −x direction (a light blue arrow) it prefers to move from one preferential direction to another (a). The direction of electric dipole moment of the 5CB molecule is indicated by the green dotted arrow. (b) shows the relative energy required to move LC molecule from one state to another. It can be seen 0.04 V/Å is enough to move the LC molecule which confirms the change of alignments of LC.